# Controlling magnetic configuration in soft-hard bilayers probed by polarized neutron reflectometry

Nan Tang[1], Jung-Wei Liao[2], Siu-Tat Chui[3], Timothy Ziman[4,5], Kai Liu[6] Chih-Huang Lai[7], Brian J. Kirby[8], Dustin A. Gilbert[1,9*]

[1] Department of Materials Science and Engineering, University of Tennessee, Knoxville, TN 37996, USA
[2] Cavendish Laboratory, University of Cambridge, Cambridge CB3 0HE, U.K
[3] Bartol Research Institute, University of Delaware, Newark, DE 19716, USA
[4] Institut Laue-Langevin, BP 156, 41 Avenue des Martyrs, 38042 Grenoble Cedex 9, France
[5] LPMMC (UMR 5493), Université de Grenobles-Alpes and CNRS, Maison des Magistères, BP 166, 38042 Grenoble Cedex 9, France
[6] Department of Physics, Georgetown University, Washington, DC 20057, USA
[7] Department of Materials Science and Engineering, National Tsing Hua University, Hsinchu, 300044, Taiwan
[8] NIST Center for Neutron Research, National Institute of Standards and Technology, 100 Bureau Dr., Gaithersburg, MD 20899, USA
[9] Department of Physics and Astronomy, University of Tennessee, Knoxville, TN 37996, USA

* Corresponding author (D.A.G.): E-mail: dagilbert@utlk.edu

**Abstract**

Hard/soft magnetic bilayer thin films have been widely used in data storage technologies and permanent magnet applications. The magnetic configuration and response to temperatures and magnetic fields in these heterostructures are considered to be highly dependent on the interfacial coupling. However, the intrinsic properties of each of the layers, such as the saturation magnetization and layer thickness, also strongly influence the magnetic configuration. Changing these parameters provides an effective method to tailor magnetic properties in composite magnets. Here, we use polarized neutron reflectometry (PNR) to experimentally probe the interfacial magnetic configurations in hard/soft bilayer thin films: $L1_0$-FePt/$A1$-FePt, [Co/Pd] /CoPd, [Co/Pt] /FeNi and $L1_0$-FePt/Fe, which all have a perpendicular magnetic anisotropy in the hard layer. These films were designed with different soft and hard layer thicknesses ($t_{soft}$ and $t_{hard}$) and saturation magnetization ($M_s^{soft}$ and $M_s^{hard}$), respectively. The influences of an in-plane magnetic field ($H_{ip}$) and temperature ($T$) are also studied using a $L1_0$-FePt/$A1$-FePt bilayer sample. Comparing the PNR results to micromagnetic simulations reveals that the interfacial magnetic configuration is highly dependent on $t_{soft}$, $M_s^{soft}$ and the external factors ($H_{ip}$ and $T$), and has a relatively weak dependence on $t_{hard}$ and $M_s^{hard}$. Key among these results, for thin $t_{soft}$, the hard and soft layers are rigidly coupled in the out-of-plane direction, then undergo a transition to relax in-plane. This transition can be delayed to larger $t_{soft}$ by decreasing $M_s^{soft}$. Understanding the influence of these parameters on the magnetic configuration is critical to designing functional composite magnets for applications.

**Introduction**

Heterostructures of coupled, magnetically hard/soft composites [1-5] have played a critical role in many modern data storage [6-17] and permanent magnet [1,18-20] technologies. In magnetic recording media, the high-anisotropy (magnetically hard) layer is frequently coupled to a magnetically soft layer which aids in the data-writing process [21-23]. In these exchange coupled composite (ECC) systems, the soft layer nucleates a domain wall that is easily pushed into the hard layer, reducing the writing field. As the anisotropy of the hard layer has been increased to allow for smaller grains, and thus higher recording densities[16,24-26], the writability provided by the soft layer has become indispensable [27,28]. In another example, high-anisotropy magnetic materials are coupled to high moment materials to develop a composite magnetic structure with a large maximum energy product for permanent magnet applications [1,29]. The magnetic configuration in composite magnets thus reflect each layer's intrinsic properties, propagated across the interface by the interface exchange and dipolar fields. Since the interfaces are highly imperfect – rich in dislocations and vacancies, becoming a virtual grain boundary – the resultant properties are an enigmatic mixture of each materials properties and can be challenging to predict. Understanding the full range of interactions within hard/soft coupled systems is critical for composite applications. Magnetic hard/soft bilayers can provide a model system to gain critical understanding of the magnetic interactions and their impacts on technological applications [30,31].

In hard/soft bilayers with perpendicularly magnetized hard layer, the magnetic interactions are highly dependent on the thickness of the soft layer, $t_{soft}$ [32,33]. For the case of a thin soft layer with a thickness less than the classical exchange length ($l_{ex}$), it is typically accepted that the magnetic moments are rigidly coupled to the hard layer by the exchange interaction. Furthermore, as $t_{soft}$ increases, it is expected that the magnetitic moment relaxes in-plane, especially away from the hard/soft interface, resulting in exchange spring-like behavior [34]. In addition to $t_{soft}$, the magnetic configuration is also influenced by the material properties including the saturation magnetization ($M_s^{hard}$ and $M_s^{soft}$), exchange stiffness ($A_{hard}$ and $A_{soft}$), the hard layer magnetocrystalline anisotropy ($K_u$) and the hard/soft interfacial coupling stiffness ($A_{int}$) [30,31,35]. While some of these magnetic material parameters have predictable relationships with the magnetic configuration, e.g. $K_u$ and $A_{hard}$ can be used to determine an exchange length ($\approx \pi\sqrt{\frac{A_{hard}}{K_u}}$) in hard magnets, the dependence on $M_s^{hard}$ and $M_s^{soft}$ or the film thickness, is not as straightforward. For modern applications in heat assisted magnetic recording (HAMR) technologies, understanding the temperature dependence of these properties and its effect on the magnetic configuration is critical to device functionality [36-40]. This quality is manifested in HAMR devices where high temperatures are used to reduce $K_u$, but simultaneously also reduce $M_s^{hard}$ and $M_s^{soft}$, establishing an optimization problem to achieve the ideal writability. Finally, the applied magnetic field also strongly influences the magnetic configuration; while most of the magnetic energy terms are uniaxial, the field dependent Zeeman energy is unidirectional and establishes a new balance between the various energy terms. Therefore, to better design and control hard/soft magnetic heterostructures, it is important to understand the roles of sample geometry ($t_{soft}$), materials properties ($M_s^{hard}$, $M_s^{soft}$) and external influences (applied in-plane magnetic field, $H_{ip}$, and temperature, $T$) in determining the magnetic configuration.

In this work, we use polarized neutron reflectometry (PNR) to investigate the magnetic configuration in hard/soft bilayer thin films. Comparing with previous magnetometry-based studies [2,30], PNR reveals the magnetic depth profile within the film, which is determined by the magnetic interactions. A series of bilayer samples with a hard magnetic underlayer (having perpendicular magnetic anisotropy) and a soft top layer is investigated, including $L1_0$-FePt(hard)/$A1$-FePt(soft), [Co/Pd](hard)/CoPd(soft), [Co/Pt](hard)/FeNi(soft) and $L1_0$-FePt(hard)/Fe(soft). By designing $M_s^{hard}$, $M_s^{soft}$ and $t_{soft}$, the magnetic interactions, and the resultant magnetic configuration, are systematically varied. Micromagnetic simulations are performed to support the experimental results and reflect similar magnetic behavior. The simulations also reveal the influence of $M_s^{hard}$ and hard layer thickness ($t_{hard}$) in determining the magnetic configuration. In addition to the intrinsic properties and sample structure, the influence of temperature ($T$) and magnetic field ($H$) on the magnetic configuration are also studied by PNR. Polarized neutron reflectometry measurements, together with micromagnetic simulations, provide a comprehensive understanding of the influence of sample structure and external control variables on the magnetic configuration of hard/soft bilayer composite magnets. These results provide insight into the underlying physics correlating bilayer film structure with magnetic configuration.

**Methods**

Four series of samples covering a range of $M_s^{soft}$, $M_s^{hard}$ and $t_{soft}$ are synthesized by magnetron sputtering and the details can be found elsewhere [32,41-43]: 1). The Si/SiO$_2$ (200 nm) /$L1_0$-FePt (4 nm)/$A1$-FePt ($t_{soft}$)/Ti (4 nm) samples were prepared in a vacuum system with a base pressure of 8×10$^{-8}$ Torr. The oriented $L1_0$-FePt was prepared by rapid thermal annealing, as described previously [26,32,44]. Without breaking vacuum, the $A1$-FePt soft layer (2nm or 5nm) was deposited on $L1_0$-FePt by co-sputtering of elemental targets, followed by a Ti capping layer. 2). The preparation process of Si/SiO$_2$ (200 nm) /$L1_0$-FePt (4 nm)/Fe (9 nm)/Pt (2 nm) is similar to the Si/SiO$_2$ (200 nm) /$L1_0$-FePt (4 nm)/$A1$-FePt ($t_{soft}$)/Ti (4 nm) samples, but sputtering 9 nm Fe and 2 nm Pt as the soft and capping layers, respectively. 3). For Si/SiO$_2$/[Co(0.4 nm)/Pt(0.7nm)]$_4$/FeNi ($t_{soft}$)/Pd(2 nm) samples, 4 repeats of Co (0.4nm)/ Pt (0.7 nm) multilayer stack were first sputtered on Si/SiO$_2$ substrate. Four different thicknesses of soft Fe$_{0.2}$Ni$_{0.8}$ (0, 1, 3, 7 nm) layers are then deposited, followed by a 2 nm Pd capping layer. 4). The Si/Pd (20 nm)/[Co (0.3 nm)/Pd (0.9 nm)]$_{25}$/CoPd($t_{soft}$)/Pd (5 nm) samples were magnetron sputtered in an ultrahigh vacuum chamber with a base pressure of 9×10$^{-9}$ Torr. Magnetically hard [Co (0.3 nm)/Pd (0.9 nm)]$_{25}$ multilayers were sputtered onto Si substrates, with a 20nm Pd buffer layer. Different thicknesses of $t_{soft}$ (2, 5, 9 nm) were then deposited, followed by a 5 nm Pd capping layer. For all the samples, the underlayer films ($L1_0$ FePt, Co/Pt and Co/Pd multilayers) have strong perpendicular magnetic anisotropy (PMA).

Polarized neutron reflectometry was performed at the NIST Center for Neutron Research on both the PBR and MAGIK instruments, which uses 4.75 Å and 5.00 Å wavelength neutrons, respectively. For the $L1_0$-FePt/$A1$-FePt samples to study the influence of $t_{soft}$ on magnetic configuration, the film was rotated into a grazing incidence with the neutron beam. The films were first saturated with a magnetic field (700 mT) parallel to the film normal direction and then the field was reduced to 50 mT. Then, the film was rotated such that the 50 mT field is applied along the film normal direction; this field is insufficient to significantly change the magnetic configuration, and acts to preserve the orientation of the neutrons, i.e. as a guide field. The out-of-plane guide field sets the orientation of the neutron spin moment always along the out-of or into-the-plane direction. The PNR measurement is executed by performing a coupled theta-2theta specular scan, with control and sensitivity to the incident and scattered neutron spin. The non-spin-flip PNR cross-sections ($R^{++}$ and $R^{--}$, where R is the reflection intensity, with the superscript indicating the spin configuration of the incident and reflected neutrons, + (–) indicting the spin-up (spin-down) geometries) typically provide sensitivity to the nuclear structure and the magnetization parallel to the neutron spin moment. However, the neutron scattering rules prohibit scattering from magnetization parallel to the momentum transfer vector, which for PNR, is in the out-of-plane direction. PNR is sensitive only to the in-plane magnetization, and with an out-of-plane guide field, the non-spin flip cross section encodes only the nuclear structure. Using an out-of-plane guide field, any net in-plane magnetization is orthogonal to the neutron spin and thus contributes to the spin-flip ($R^{+-}$, $R^{-+}$) cross section. In this geometry, sensitivity to the orientation of the in-plane magnetization is lost, however the signal is increased by including all of the magnetic signal into the spin-flip cross section.[45,46] For all other PNR measurements, the samples were saturated in the out-of-plane direction, then returned to remanence and measured in a small in-plane guide field (1 mT). In this configuration, $R^{++}$ and $R^{--}$, provide sensitivity to both the nuclear depth profile, and the depth profile of the net in-plane magnetization component parallel to the applied field. Spin-flip scattering for this configuration is not considered here. All the experimental data are then model fitting by REFL1D software [45].

Three dimensional micromagnetic simulations are performed using the object oriented micromagnetic framework (OOMMF) platform,[47] with a simulation volume of 200 nm × 1 nm × ($t_{hard} + t_{soft}$) and a mesh of 0.5 nm × 0.5 nm × 0.2 nm. The exchange stiffness for hard magnet ($A_h$), hard/soft interface ($A_{int}$), and soft magnetic ($A_s$) are $A_h = A_s = A_{int} = 1.15 \times 10^{-11}$ J/m. The uniaxial anisotropy for hard magnet ($K_u$) is chosen to be $1 \times 10^6$ J/m$^3$ while the anisotropy for soft magnet ($K_s$) is 0.

**Experimental results**

*Effect of soft layer thickness*

With the above experimental preparations, the first set of measurements were performed to study the influence of $t_{soft}$ on the magnetic configuration. Using an out-of-plane guide field, reflectometry data for the $L1_0$-FePt(4 nm)/$A1$-FePt($t_{soft}$) bilayer films, with $t_{soft} = 2$ nm and 5 nm, is shown in Figure 1a and b, respectively. Data on the same sample, but taken at a different time, was shown in reference [32], and reflects a similar profile; differences in the capping layer structure are captured in this work, likely due to the higher q-range measured. For the $t_{soft}=2$ nm film, the experimental non-spin-flip $R^{++}$ and $R^{--}$ are essentially identical, and the spin-flip $R^{+-}$ and $R^{-+}$ are virtually zero. This indicates that there is little net in-plane magnetization anywhere in the film. In comparison, for the $t_{soft}=5$ nm film, the $R^{++}$ and $R^{--}$ are still overlapping, but $R^{+-}$ and $R^{-+}$ are now non-zero, indicating the presence of appreciable in-plane magnetization. These interpretations are confirmed by model fitting (solid lines in Figure 1a and 1b) that yields the scattering length density (SLD) [48] depth profiles for the two films, shown in Figure 1c and d. The profiles have a nuclear component (NSLD) indicative of the nuclear composition, and a magnetic component (MSLD) proportional to the in-plane magnetization. Notably, the MSLD is approximately zero for $t_{soft}$=2 nm, and non-zero for $t_{soft}$=5 nm. These results confirm that $t_{soft}$ can influence the magnetic configuration in the soft layer. Interestingly, this influence extends all the way to the interface: while both the hard and soft layers in the 2nm sample are oriented out-of-plane, the soft layer in the 5nm sample is almost entirely in-plane.

*Effect of saturation magnetization*

Besides $t_{soft}$, $M_s^{hard}$ and $M_s^{soft}$ are expected to influence the magnetic configurations in these bilayer films. PNR measurements are also performed on a series of samples with $M_s^{hard} \ll M_s^{soft}$, $M_s^{hard} < M_s^{soft}$ and $M_s^{hard} > M_s^{soft}$, including $L1_0$-FePt(4nm)/Fe(9nm), [Co (0.3 nm)/Pd(0.9 nm)]$_{25}$/CoPd(9nm), and [Co(0.4 nm)/Pt(0.7nm)]$_4$/FeNi (7 nm), respectively. For the case of $M_s^{hard} \ll M_s^{soft}$ shown in Figure 2a, the experimental R$^{++}$ and R$^{--}$ curves are significantly separated, which indicates a net in-plane magnetization. The corresponding magnetic depth profiles, in Figure 2b, show a large MSLD in the soft layer (Fe region). Using the MSLD for bulk iron (4.85 ×10$^{-4}$ nm$^{-2}$) [48], the soft Fe layer is confirmed to be almost entirely in-plane throughout its thickness. For $M_s^{hard} < M_s^{soft}$ and $M_s^{hard} > M_s^{soft}$, the experimental R$^{++}$ and R$^{--}$ curves are strongly overlapping, as shown in Figures 2c and e, respectively. The corresponding magnetic depth profiles show almost no MSLD in Figures 2d and f, respectively. These results suggest that in both the [Co (0.3 nm)/Pd (0.9 nm)]$_{25}$/CoPd (9 nm) and [Co (0.4 nm)/Pt (0.7nm)]$_4$/FeNi (7 nm) films, even at these relatively larger thicknesses, the soft layer magnetization remains mostly in the out-of-plane direction. The PNR measurements are also performed on another five [Co (0.3 nm)/Pd(0.9 nm)]$_{25}$/CoPd and [Co(0.4 nm)/Pt(0.7nm)]$_4$/FeNi samples with thinner soft layer thickness, shown in S1. These films also show very little in-plane magnetization.

*Effect of in-plane field and temperature*

The results above implicate a complex competition within these systems to define the soft-layer orientation, which depends on both $M_s^{soft}$ and $t_{soft}$. Besides these intrinsic material parameters, the interfacial magnetic configuration is also dependent on the external variables ($H_{ip}$ and $T$). Six PNR measurements were performed on the $L1_0$-FePt/$A1$-FePt (2 nm) bilayer sample with $\mu_0 H_{ip}$ from 100 mT to 700 mT at T=295 K and T=500 K, shown in Figure 3a and b, respectively. For each in-plane magnetic field, the corresponding R$^{++}$ and R$^{--}$ curves are separated, which indicates a net in-plane magnetization. This result can be alternatively confirmed by the spin asymmetry (SA) shown in Figure 3c and d for $T$=295 K and $T$=500 K, respectively. SA is a quantity expressed as $\frac{R^{++}-R^{--}}{R^{++}+R^{--}}$, which highlights magnetic contributions to the scattering. Both Figure 3c and d show a non-zero SA feature at all $H_{ip}$ and $T$, indicating the presence of a net in-plane magnetization for all the measurements. Fitted by REFL1D, the nuclear and magnetic depth profiles determined from model fitting are shown in Figure 3e and f, respectively. The fitting constrained the nuclear depth profiles to be the same for all the measurements since the nuclear structure was not expected to change with $T$ and $H_{ip}$. The magnetic depth profiles here are presented using both MSLD (left axis) and the converted in-plane magnetization (right axis). All the PNR measurements show an appreciable in-plane component of the magnetization. This relaxation follows a power law fall off from the maximal magnetization in the soft layer ($M^{top}$) to the in-plane component of the magnetization of hard underlayer ($M^{bot}$). As an example, the position of $M^{top}$ and $M^{bot}$ for $\mu_0 H_{ip}$=100 mT and T=295 K is demonstrated (dotted blue line). At both $T$=295 K and $T$= 500 K, increasing $H_{ip}$ results an overall increase of the in-plane magnetization, including $M^{top}$ and $M^{bot}$, which can be attributed to the increase of in-plane Zeeman energy.

An alternative analysis of the magnetic relaxation is performed by normalizing the magnetic depth profiles by their respective $M^{top}$, shown in Figure 3g. For each $H_{ip}$, the maximum normalized magnetization (M/$M^{top}$=1) appears at almost the same position in the soft layer, about 5.8 nm away from the SiO$_2$ interface. Notably, increasing $H_{ip}$ causes an increase of the interfacial domain wall width, $l_{int}$,[49] particularly by extending further into the hard layer. The $H_{ip}$ presents an important role in determining the interfacial magnetic configurations by inducing an external Zeeman energy, while increasing $T$ enhances thermal fluctuations in the bilayer, reducing the anisotropy and competing with the exchange coupling. At each $H_{ip}$, increasing the temperature results in an overall reduction of the total magnetization including $M^{top}$ and $M^{bot}$. For $\mu_0 H_{ip}$=100 mT, $M^{top}$ changes from 320 kA/m at 295 K to 180 kA/m at 500 K, more than 40% reduction. This temperature driven reduction of $M^{top}$ decreases with increasing $H_{ip}$, becoming 30% for $\mu_0 H_{ip}$= 300 mT and only 16% for $\mu_0 H_{ip}$= 700 mT. These results indicate that increasing $H_{ip}$ can compensate the thermal fluctuations, therefore increasing the in-plane soft layer magnetization. The normalized magnetic depth profiles for each $H_{ip}$ shows that the interfacial magnetic configuration at 500 K is similar to the configuration at 295 K, indicating that there is almost no change in $l_{int}$ or the normalized relaxation rate with temperature between 295 K and 500 K. These PNR experimental results reveal the influence of $M_s^{soft}$, $t_{soft}$, $H_{ip}$ and $T$ on the interfacial magnetic configuration, summarized in Table 1.

**Discussion and micromagnetic simulation**

The PNR results provide a direct measurement of the magnetic relaxation within the bilayer system; the simulated magnetic depth profiles suggest this relaxation follows a power law from $M^{top}$ to $M^{bot}$. PNR measurements show that this magnetic relaxation is dependent on sample geometry ($t_{soft}$), material properties ($M_s^{soft}$) and external factors ($H_{ip}$ and $T$). For $t_{soft}$ larger than a critical thickness, the magnetic configuration changes from a rigid coupling, in which the soft layer has an out-of-plane orientation reflecting the hard underlayer, to an in-plane relaxation, in which the soft layer rotates from out-of-plane to in-plane moving from the interface; this critical soft layer thickness increases with decreasing $M_s^{soft}$. Increasing $H_{ip}$ is shown to increase the overall in plane magnetization and $l_{int}$, while increasing $T$ from room temperature to 500 K result in a reduction of the in-plane magnetization, but $l_{int}$ remains almost unchanged. In a simplified model, the in-plane relaxation can be considered as a balance among the magnetostatic energy of the soft layer, perpendicular anisotropy of the hard layer, the interfacial exchange coupling between the soft and hard layers and the Zeeman interaction. The increase of $t_{soft}$ or $M_s^{soft}$ results in the increase of total demagnetization energy, as well as the decrease of energy gap between the ground state to the excited magnon state [50,51]. When the thermal energy ($k_B T$) is high enough to overcome this energy gap, it will cause the transition from rigid coupling to the underlayer, which favors an in-plane orientation for the soft layer magnetization. As a result, the interfacial magnetic configuration relaxes in-plane. For increasing $H_{ip}$, the Zeeman energy favors an in-plane orientation parallel to $H_{ip}$, which increases $M^{top}$ and $M^{bot}$. However, while increasing $H_{ip}$, the in-plane magnetic component of the soft layer increases more than the hard layer, due to the much larger saturation magnetization and the absence of a perpendicular anisotropy. The angle between the magnetization vector of the bottom hard layer and that of the top of the soft layer increases. This increased angle causes the interfacial exchange coupling energy to increase commensurately. To compensate this increase of energy, $l_{int}$ extends further into the hard layer. Increasing $T$ to 500 K, the thermal fluctuations are enhanced which decreases $M_s^{soft}$, $M_s^{hard}$ and $K_u$. The reduction of $M_s^{soft}$ decreases $l_{int}$, while the reduced $K_u$ should increase $l_{int}$; these two factors can compensate each other, resulting an almost unchanged $l_{int}$. Indeed, in the experimental results $l_{int}$ is almost unchanged. Ideally, increasing $l_{int}$ improves writability in HAMR media[52], suggesting that, in this bilayer system, reducing the temperature dependence of $M_s^{soft}$ by e.g. increasing the Curie temperature, would be a promising direction for improvement. The PNR measurements reveal the important roles of $t_{soft}$, $M_s^{soft}$, $H_{ip}$ and $T$ in determining the configuration of the magnetic bilayer. The following simulations reinforce these observations and provide a more complete picture which must also include $t_{hard}$ and $M_s^{hard}$.

The first set of simulations that are presented explore the case of $M_s^{soft}$ (300 kA/m) $<M_s^{hard}$ (900 kA/m), over a range of $t_{soft}$ similar to our experiments with [Co/Pt]/NiFe, shown in Figure 4a. For thin $t_{soft}$ (2 nm and 5 nm), the magnetic moments in the soft layer are strictly orientated in the out-of-plane direction. This magnetic configuration can be understood by the classical exchange length, $l_{ex}$: that for $t_{soft} < l_{ex}$ the interfacial exchange coupling dominates the energy landscape, and rigidly couples the hard and soft layers. Increasing $t_{soft}$ to 9 nm, the in-plane magnetic component of the soft layer begins to increase. Interestingly, the relaxation in the soft layer extends all the way to the interface, challenging the initial idea that the hard/soft exchange coupling is dominant in the interfacial region. Further increasing $t_{soft}$ to 20 nm and 30 nm, the relaxation rate of magnetic configuration in the soft layer decreases, e.g. the soft layer recovers an out-of-plane orientation at the hard/soft interface. These results show a complex and nonmonotonic dependence of the soft-layer orientation on $t_{soft}$. For completeness, the dependence of the magnetic relaxation on $t_{hard}$ was also investigated in the case of $M_s^{soft}=M_s^{hard}=700$ kA/m, shown in Figure 4b. Increasing $t_{hard}$ does not significantly impact the magnetic configuration in the soft layer, indicating that the interfacial exchange interaction dominates the interaction between hard and soft layers.

These results show a $t_{soft}$ dependent and non-$t_{hard}$ related magnetic relaxation: for $t_{soft}<l_{ex}$, the hard and soft layers are rigidly exchange coupled, while for thicker $t_{soft}$, the soft layer relaxes in-plane with a relaxation rate which decreases as increasing $t_{soft}$. The next set of simulations explore the role of $M_s^{soft}$ and $M_s^{hard}$ in determining the magnetic bilayer configuration while fixing $t_{soft}$ and $t_{hard}$. For a fixed $M_s^{hard}=700$ kA/m, $M_s^{soft}$ is varied between (300-1700) kA/m, and the out-of-plane magnetization is shown in Figure 4c. Increasing $M_s^{soft}$ significantly accelerates the in-plane relaxation of the magnetic soft layer. This acceleration can be seen by evaluating the distance from the hard/soft interface at-which the soft layer relaxes in-plane. For $M_s^{soft}=300$ kA/m, this distance is >30 nm, while for $M_s^{soft}=1700$ kA/m this distance is only 2 nm. In contrast, models with a fixed $M_s^{soft}$ and variable $M_s^{hard}$ between 300 kA/m and 1300 kA/m, Figure 4d, show comparably little change. The bilayer simulations show generally that increasing $M_s^{hard}$ causes the in-plane relaxation to occur over a larger thickness, e.g. the out-of-plane component of the soft-layer magnetization extends further from the hard/soft interface.

These simulations agree with the experimental results. For example, in the FePt/Fe(9 nm) sample the soft-layer was almost fully in-plane, in strong contrast with the [Co/Pd]/CoPd(9 nm) sample which was about fully out-of-plane. The simulations would suggest that the differences in the hard layer have little effect, while the $M_s^{soft}$ dominates the magnetic configuration. Indeed, with a much-larger $M_s^{soft}$, Fe ($M_s^{soft}$=1700 kA/m) is expected to be in-plane, while CoPd ($M_s^{soft} \approx 750$ kA/m) could remain out of plane.

**Conclusion**

In summary, we have investigated magnetic configurations in perpendicularly-coupled hard/soft bilayer films as a function of their geometric structure ($t_{soft}$), material properties ($M_s^{soft}$ and $M_s^{hard}$) and external influences ($H_{ip}$ and $T$). Using polarized neutron reflectometry, the depth profiles for these coupled systems are experimentally determined and compared to micromagnetic simulations. For very thin $t_{soft}$, on the order of $l_{ex}$, the magnetic moments in the soft layer are rigidly coupled to the hard layer throughout their entire thickness, while for thicker $t_{soft}$ the magnetic moments in the soft layer tend to relax in plane. Surprisingly, the in-plane relaxation extends into regions which were oriented in the out-of-plane direction for smaller $t_{soft}$, indicating the relaxation is not simply progressive with increasing $t_{soft}$. The in-plane relaxation is also shown to be dependent on $M_s^{soft}$, with increasing $M_s^{soft}$ significantly increasing its rate. These results are directly demonstrated in [Co/Pd]/CoPd, [Co/Pt]/FeNi, $L1_0$-FePt/Fe, and $L1_0$-FePt/$A1$-FePt bilayer films using PNR, then confirmed by micromagnetic simulations. Moreover, the simulation results suggest that increasing $M_s^{hard}$ decreases the relaxation rate in the soft layer, while changes in $t_{hard}$ may not significantly influence the magnetic configuration. PNR experiments also reveal the importance of $H_{ip}$ and $T$ on the magnetic configuration. Specifically for the $L1_0$-FePt/$A1$-FePt system, increasing $H_{ip}$ or decreasing $T$ results in an overall increase of the in-plane magnetization. This result corresponds to the temperature-dependent balance of the magnetocrystalline anisotropy and the magnetostatic energy. Meanwhile, increasing $H_{ip}$ causes the measured $l_{int}$ to increase, while increasing $T$ leaves $l_{int}$ almost unchanged. The PNR and simulation results together show that the magnetic configuration in the bilayer samples can be modulated by varying $t_{soft}$, $M_s^{hard}$ and $M_s^{soft}$ as well as $H_{ip}$ and $T$. Understanding the impact of these parameters is critical for the design of hard/soft bilayer devices.

**Acknowledgement**

This material is based upon work supported by the U.S. Department of Energy, Office of Science, Office of Basic Energy Sciences Energy Early Career Research Program under Award Number DE-SC0021344. Work at G.U. has been supported by the NSF (ECCS- 1933527).

**Supplementary materials**

PNR measurements are also performed on [Co/Pd]-multilayers/CoPd with $t_{soft}$= 2 nm and 5 nm. The samples are first saturated with 500 mT magnetic field along film normal direction, and then reduce to 0. An in-plane guide field 1 mT was applied while measuring. The reflectivity of R$^{++}$ and R$^{--}$ are largely overlapped with each other for $t_{soft}$=2 (Figure S1 a and b), which indicates there is not net in plane magnetization. This is then proved by converged magnetic depth profiles that no net MSLD can be probed. This non-in-plane magnetic behavior is also probed in $t_{soft}$=5 nm (Figure S1 c and d).

With the same experimental set up on the above, PNR measurements and the model fitting on [Co/Pt]/FeNi with $t_{soft}$=0, 1 and 3 nm are shown in Figure S2 (a, b), (c,d) and (e,f), respectively. For each $t_{soft}$, R$^{++}$ and R$^{--}$ are largely converged with each other and the corresponding magnetic depth profile presents no MSLD. These results indicate that the magnetization in the soft layer is along the perpendicular anisotropy orientation.

**Figures**

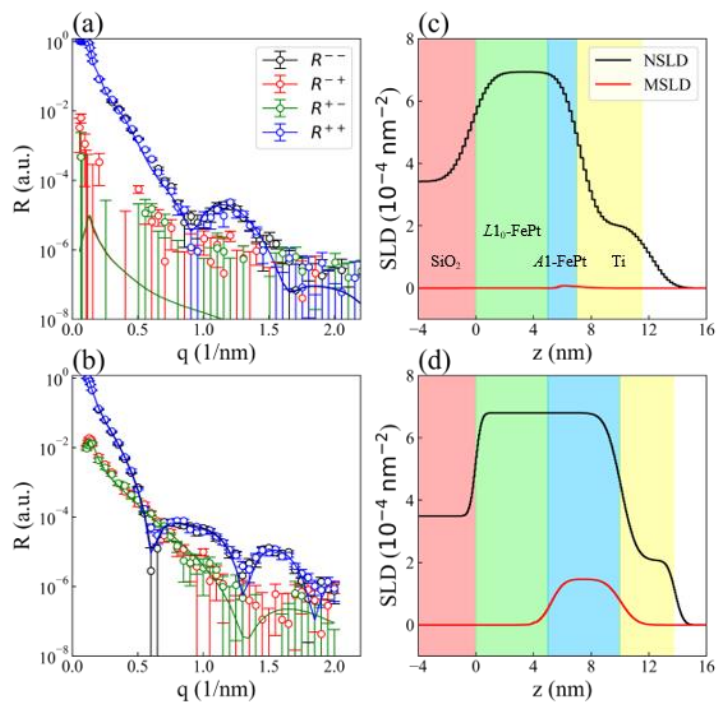

**Figure 1** Model-fitted PNR data of $L1_0$-FePt (4nm)/$A1$-FePt for (a) $t_{soft}$=2 (b) $t_{soft}$=5 nm and the associated scattering length density profiles (c) $t_{soft}$=2 (d) $t_{soft}$=5 nm. The red, green, blue and yellow regions correspond to the depth profile of $SiO_2$, $L1_0$-FePt, $A1$-FePt and capping Ti layers, respectively.

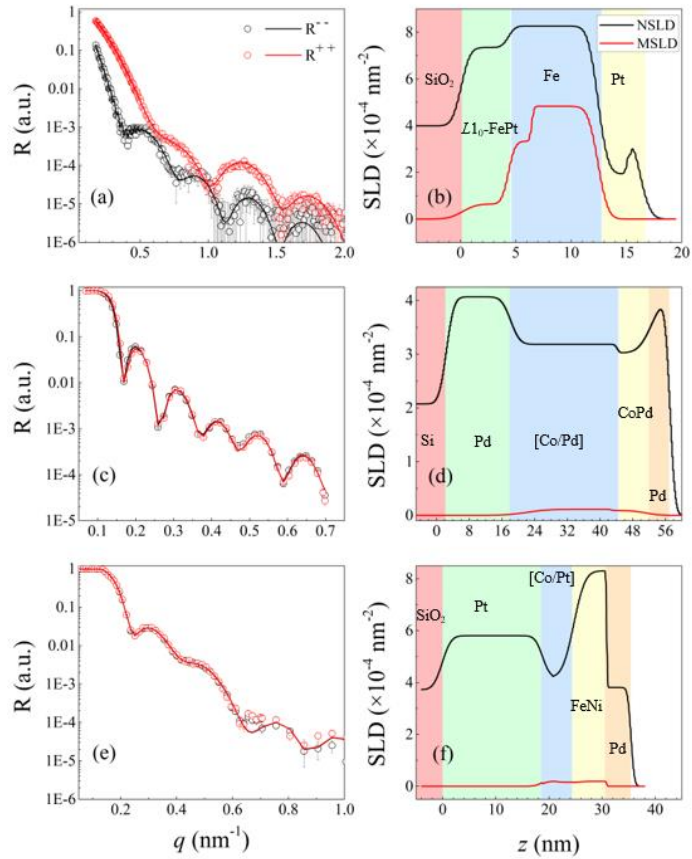

**Figure 2** $L1_0$-FePt/Fe (a) Fitted $R^{++}$ and $R^{--}$ and (b) the corresponding nuclear (black) and magnetic (red) SLD profiles, where pink, green, blue and yellow regions are the $SiO_2$, $L1_0$-FePt, Fe and Pt capping layers, respectively. [Co/Pd]-multilayers/CoPd (c) Fitted $R^{++}$ and $R^{--}$ and (d) nuclear (black) and magnetic (red) SLD profiles where pink, green, blue, yellow and orange regions correspond to the Si, Pd seed layer, [Co/Pd] multilayers, CoPd soft layer and Pd capping layer, respectively. [Co/Pt]-multilayers/FeNi (e) fitted $R^{++}$ and $R^{--}$ and corresponding nuclear (black) and magnetic (red) SLD profiles with $SiO_2$ (pink), Pt buffer layer (green), [Co/Pt] multilayers (blue), FeNi (yellow) and Pd capping layer (orange).

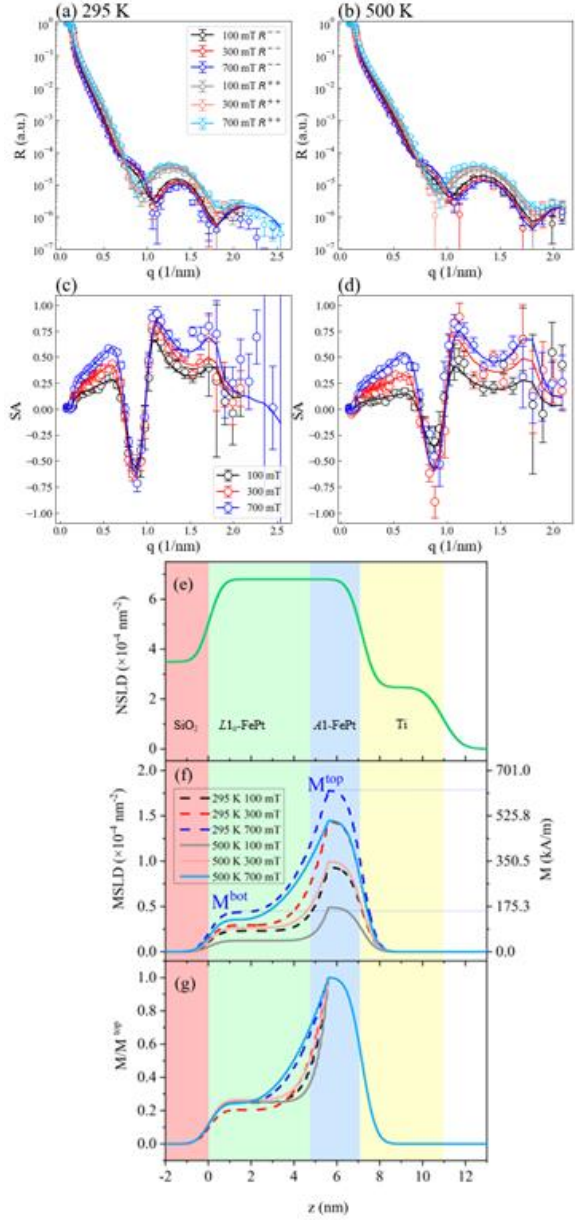

**Figure 3** Polarized neutron reflectivity for $L1_0$-FePt (4 nm)/$A1$-FePt (2nm) with in-plane magnetic fields of 100 mT, 300 mT and 700 mT at (a) 295 K and (b) 500 K. The spin asymmetry for each field at (c) 295 K and (d) 500 K. The fitted (e) nuclear and (f) magnetic depth profiles and (g) the normalized magnetic depth profile. $M^{top}$ corresponds to the maximal in-plane magnetization of the soft layer and $M^{bot}$ is the magnetization of the hard layer.

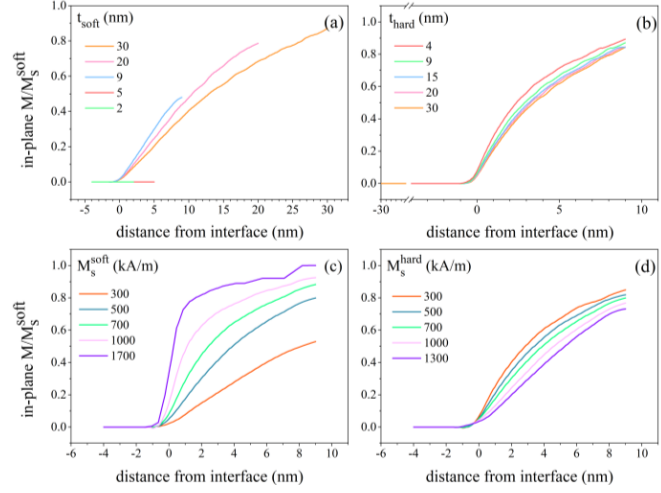

**Figure 4** OOMMF simulation for the out-of-plane magnetization for (a) $M_S^{hard}$=900 kA/m and $M_S^{soft}$=300 kA/m and $t_{hard}$=4 nm with $t_{soft}$ = (2, 5, 9, 20, 30) nm (b) $M_S^{hard}$=700 kA/m and $M_S^{soft}$=700 kA/m and $t_{soft}$=9 nm with $t_{hard}$ = (4, 9, 15, 20, 30) nm, (c) $M_S^{hard}$=700 kA/m and $M_S^{soft}$= (300, 500, 700, 1000, 1700) kA/m (d) $M_S^{soft}$=500 kA/m and $M_S^{hard}$=(300, 500, 700, 1000, 1300) kA/m, where $t_{hard}$=4 nm and $t_{soft}$ =9 nm

Table 1: the summary of PNR experimental results

| Sample | $T_{soft}$ (nm) | Soft-layer MSLD ($\times 10^{-4}$ nm$^{-2}$) | Guide Field (mT) | Temperature (K) |
|---|---|---|---|---|
| $L1_0$-FePt/A1-FePt | 2 | 0.07 | 50 (out-of-plane) | 295 |
| | 5 | 1.46 | 50 (out-of-plane) | 295 |
| | 2 | 0.93 | 100 | 295 |
| | 2 | 1.41 | 300 | 295 |
| | 2 | 1.77 | 700 | 295 |
| | 2 | 0.49 | 1 | 500 |
| | 2 | 0.99 | 1 | 500 |
| | 2 | 1.44 | 1 | 500 |
| [Co/Pd]/CoPd | 2 | 0.01 | 1 | 295 |
| | 5 | 0.04 | 1 | 295 |
| | 9 | 0.08 | 1 | 295 |
| [Co/Pt]/FeNi | 1 | 0.37 | 1 | 295 |
| | 3 | 0.22 | 1 | 295 |
| | 7 | 0.19 | 1 | 295 |
| $L1_0$-FePt/Fe | 9 | 4.82 | 1 | 295 |